\documentclass[a4paper,11pt]{article}
\usepackage{jheppub} 
\usepackage{lineno}
\usepackage{orcidlink}
\usepackage{tikz}
\usepackage{tkz-euclide}
\usetikzlibrary{shapes}	
\tikzset{every picture/.style={line width=1}} 
\usepackage{indentfirst}
\arxivnumber{2508.03797}
\title{Cosmic Axion Background Detection Using Resonant Cavity Arrays}
\newcounter{qnumber}
\newcounter{pnumber}

\renewcommand{\sc}[1]{\stepcounter{pnumber}{\color{red} [SC~\arabic{pnumber}:\,#1]}}

\author[]{Soobeom Chung\,\orcidlink{0009-0006-7731-5395} and Jeff A. Dror\,\orcidlink{0000-0003-0110-6184}}
\affiliation{Institute for Fundamental Theory, Physics Department, University of Florida, Gainesville, FL 32611, USA}
    
\emailAdd{soobeomchung@ufl.edu}
\emailAdd{jeffdror@ufl.edu}

\abstract{The axion is a well-motivated and generic extension of the Standard Model. If produced in the early universe, axions may still be relativistic today, forming a Cosmic Axion Background (C$a$B) potentially detectable in direct detection experiments. Although C$a$B is expected to be broadband, which makes it challenging to be detected, a high-quality-factor microwave cavity acts as a narrowband filter with response peaked at its resonant frequency. We propose a new strategy using multi-cavity arrays to distinguish signal from background noise by exploiting spatial correlations of the axion-induced electric field which are set by the cavity quality factor. We compute the two-point correlation function for electric fields in spatially separated cavities sourced by an isotropic C$a$B. Analyzing various cavity geometries, we find that stacked, wide-base cavity arrays offer coherent enhancement of the axion signal. We apply our formalism to prospective upgrades of the ADMX experiment, including configurations with four and eighteen coupled cavities. Although these arrays do not achieve a coherent enhancement, optimizing the geometry could potentially yield an $\mathcal{O}(1)$ improvement in the sensitivity to the C$a$B.}

\begin{document}
\maketitle
\flushbottom

\section{Introduction}
\label{sec:intro}
Axions are simple and compelling extensions of the Standard Model, capable of addressing the strong CP problem~\cite{Peccei:1977hh, Peccei:1977ur,Weinberg:1977ma,Wilczek:1977pj} and serving as viable dark matter candidates~\cite{Abbott:1982af, Dine:1982ah, Preskill:1982cy}. Over the past few decades, their detection has become a global effort, with different searches targeting a range of axion masses and interaction types. The most established method to search for axions from the early universe interacting with photons is the use of resonant microwave cavities, as implemented by ADMX~\cite{ADMX:2001dbg,ADMX:2009iij,Shokair:2014rna,ADMX:2018gho,ADMX:2019uok,ADMX:2025vom}, HAYSTAC~\cite{AlKenany:2016trt,Brubaker:2016ktl,Brubaker:2017rna}, and CAPP~\cite{Youn:2024omj}. 

In addition to potentially constituting dark matter, axions are well-motivated candidates for particles produced in the early universe that remain relativistic today. They could arise through thermal collisions with Standard Model particles, parametric resonance from displaced fields during inflation, decays of topological defects formed during a cosmic phase transition, or the decay of dark matter itself~\cite{Dror:2021nyr}. A relic population of such axions would form a {\em Cosmic Axion Background} (C$a$B), a classical field analogous to dark matter. The key distinction lies in the signal bandwidths of these two components.

For axion dark matter, the signal power is sharply peaked around the axion mass $m_a$. In particular, for incoming axions with speed $v$, the peak power is at an oscillation frequency $\omega \simeq m_a(1+\frac{v^2}{2})$. Since locally $v\sim 10^{-3} $, the energy distribution of dark matter is narrow, with a relative bandwidth of $v^2 \sim 10^{-6}$. In contrast, the C$a$B typically exhibits a much broader energy distribution, with a relative bandwidth of order unity, leading to a significantly shorter coherence time and length. 

 In the presence of a C$a$B, resonant microwave cavities immersed in a strong magnetic field can generate an electromagnetic response via axion-photon coupling, governed by the interaction term $g_{a\gamma\gamma}a\mathbf{E}\cdot {\bf B}$, where $g_{a\gamma\gamma}$ is the axion-photon coupling constant. The ADMX collaboration searched for a C$a$B arising from dark matter decay using existing data~\cite{ADMX:2023rsk}. Since dark matter in the Milky Way is concentrated near the Galactic Center, the resulting axion flux has a preferred direction, leading to an anisotropic C$a$B. As the orientation of the experimental apparatus changes over the course of the Earth's rotation, the cavity form factor varies, producing a characteristic modulation of the signal. This modulation served as an effective means of distinguishing signal from background noise. However, the method is only applicable to anisotropic signals and was further limited by significant noise gain in the microwave electronics. 
 
 In this study, we explore a multi-cavity array --- a key feature in future ADMX upgrades~\cite{Yang:2020xsc, Knirck2023} --- in which multiple spatially separated cavities, with dimension comparable to the coherence length of C$a$B, are used to correlate axion-induced signals. A resonant cavity with quality factor $Q$ responds as a narrowband filter. Even if the raw C$a$B field would have looked incoherent, the filtered field allows the use of inter-cavity correlations, eliminating the need to rely on anisotropy. This approach provides additional information beyond that accessible to a single detector. We extend the likelihood framework to incorporate multi-cavity setups and investigate their potential to enhance axion searches. We explore the cavity geometries that maximize the correlation signal as well as the configurations proposed for ADMX upgrades. 

 The outline of this paper is as follows. In Section~\ref{sec:2}, we derive the signal generated by an axion field and its associated correlation function in terms of the position of the cavities. We show that the function depends on the relative displacement, not on their individual locations. Then, we compute the relativistic form factor in Section~\ref{sect: correlation function} and briefly look at the statistics of a multi-cavity array using the likelihood function in Section~\ref{The statistics of a multi-cavity array}. Finally, in Section~\ref{sec:5 evaluating upper limit}, experimental sensitivity of different arrays is investigated for an ideal configuration.

\section{Resonant Cavity Correlation Function}
\label{sec:2}
We begin by reviewing the formulation of the axion field in the classical limit, following the formalism of Ref.~\cite{Dror:2021nyr, Foster:2017hbq, Foster:2020fln}. Starting from the phase-space distribution, we derive the electric field generated by the modified Maxwell equations. We then compute its correlation function across spatially separated cavities. This leads to an expression for the relativistic form factor, which captures the dependence of the multi-cavity array's geometry on the signal response.  

\subsection{Resonant Cavity Detection}
\label{sec:2.2}
 A relativistic axion, like its non-relativistic counterpart, can be described as a superposition of plane waves \cite{Dror:2021nyr}:
\begin{align}
     a(t, \mathbf x)=\sum_{\mathbf k}\sqrt{\frac{f(\mathbf{k})d^3k}{(2\pi)^3\omega_{\mathbf k}}}\alpha_{\mathbf k}\cos{(\omega_{\mathbf k}t-\mathbf{k}\cdot \mathbf{x}+\phi_{\mathbf k})}\,,
\end{align}
which is expressed in terms of a Rayleigh distributed amplitude $\alpha_{\mathbf k}$ and a single uniformly sampled phase $\phi_{\mathbf k}$.

At a single point in space, the $\mathbf{k} \cdot {\bf x}$ term can often be absorbed into the phase, up to corrections arising from spatial gradients in the axion field. However, for multi-cavity axion detection experiments, the axion field must be considered over relatively large spatial separations. This spatial extent reduces the coherence of the axion field across widely separated regions, which in turn suppresses constructive interference.

 Resonant cavity axion experiments employ low-temperature microwave cavities. A cavity mode is resonantly excited when the axion frequency matches a cavity resonant frequency $\omega_0$, which is determined by the cavity's dimensions. Typical experimental setups feature cavities of size ${\cal O}(1-100~{\rm cm})$, corresponding to axion frequencies in the $1-100~\mu {\rm eV} $ range.

If axions couple to photons, they modify the electromagnetic Lagrangian by adding an interaction term $g_{a\gamma\gamma}a\mathbf{E}\cdot {\bf B}$~\cite{Sikivie:1983ip}. The dominant effect of this term is to alter Maxwell's equations, introducing new interactions between the axion and electromagnetic fields. The influence of the axion can be treated perturbatively. We expand the fields as: 
\begin{align}
{\bf E} &= {\bf E}_0 + {\bf E}_a \,,\\
{\bf B} &= {\bf B}_0 + {\bf B}_a\,,
\end{align}
where ${\bf E}_0$ and ${\bf B}_0$ are the zeroth-order fields and ${\bf E}_a$ and ${\bf B}_a$ are their first-order corrections proportional to the axion field. Substituting this expansion into Maxwell's equations, we obtain:
\begin{align}
    \nabla\cdot\mathbf{E}_a&=-g_{a\gamma\gamma}{\mathbf B}_0\cdot\nabla a\,,\\
    \nabla\times\mathbf B_a&=g_{a\gamma\gamma}(\mathbf B_0\dot a-\mathbf E_0\times\nabla a)\,.
\end{align}

In the presence of a strong static magnetic field, a C$a$B induces oscillating electromagnetic fields in the cavity. If the static background field is uniform in space, the induced electric field satisfies: 
\begin{align}
\label{maxwell}
    (\partial_t^2-\nabla^2)\mathbf E_a=-g_{a\gamma\gamma}\mathbf B_0\ddot a\,.
\end{align}
Here, we have omitted a term proportional to the effective charge density $g_{a \gamma \gamma} {\bf B}_0 \cdot \nabla a $, which does not contribute to excitation of resonant cavity modes of interest~\cite{Dror:2021nyr}. 

Eq.~\eqref{maxwell} constitutes a sourced wave equation, which we solve inside a cylindrical cavity in the presence of a static magnetic field applied along the cavity's symmetry axis, which we take to be along $\hat{{\bf z}}$. The cavity supports a discrete set of eigenmodes $\mathbf e_{\ell mn}$ with eigenfrequencies $\omega_{\ell mn}$, labeled by three integers $(\ell, m ,n)$, satisfying 
\begin{align}
    (\nabla^2+\omega^2_{\ell mn})\mathbf e_{\ell mn}(\mathbf x)&=0\,.
    \end{align}
These eigenmodes are normalized according to:
\begin{align}
    \int d^3x ~\mathbf e_{\ell mn}\cdot{\mathbf e}^*_{\ell'm'n'}&=\delta_{\ell \ell '}\delta_{mm'}\delta_{nn'}\,,
\end{align}
where the integral covers the cavity volume $V$. These modes form a complete basis for expanding the electric field within the cavity. 

To solve Eq.~\eqref{maxwell}, we expand the induced electric field in this eigenmode basis:  
\begin{equation}
\label{eigenmode}
    \mathbf{E}_a(t,{\mathbf x})=\sum_{\ell mn}\alpha_{\ell mn}(t)\mathbf e_{\ell mn}({\bf x})\,.
\end{equation}
We define the temporal Fourier transform of the electric field coefficient over the duration of the experiment $T$: 
\begin{align}
    \tilde\alpha_{\ell m n} (\omega)=\frac{1}{\sqrt T}\int^{T/2}_{-T/2} dt~\alpha_{\ell m n }(t)e^{-iwt}\,.
\end{align} 
To include dissipative effects, we introduce a $-\omega_{\ell mn}\partial_t\mathbf E_a/Q$ term to the LHS of Eq.~\eqref{maxwell}~\cite{Brubaker:2017ohw}, where $Q$ is the quality factor of the cavity. Substituting Eq.~\eqref{eigenmode} into the wave equation, we find:
\begin{align}
    \label{EFC frequency}
\tilde\alpha_{\ell mn}(\omega)&=-g_{a\gamma\gamma}B_0{\cal T}(\omega)\int d^3x\tilde a(\omega,  {\mathbf x})\mathbf{e}^*_{\ell mn}({\bf x})\cdot\mathbf{\hat{z}}\,.
\end{align}
Here we introduced the notation, ${\cal T}(\omega) =\omega^2(\omega^2-\omega^2_0-i\omega\omega_{0}/Q)^{-1}$, to denote the cavity's frequency-dependent transfer function \cite{Dror:2021nyr}. 

 We focus on the mode most commonly monitored in axion detection experiments: the transverse magnetic mode $\text{TM}_{100}$. Its eigenfunction is
\begin{align}
\label{eigenfunction}
    (\mathbf{ e}_{100}({\bf x}))_z= \frac{1}{\sqrt V}\frac{J_0(\omega_0r)}{J_1(\omega_0R)}\equiv (\mathbf{ e}({\bf x}))_z\,,
\end{align} 
where $J_n$ are Bessel functions of the first kind, $r$ is the radial distance from the cavity axis, $R$ is the cavity radius, and $\omega_0 = j_{10}/R$ is the resonant frequency, with $j_{10}$ the first zero of $J_0$. For brevity, we omit the mode indices moving forward.

\subsection{Correlation Function}
\label{sec2.3}
By explicitly computing the Fourier transformed field, we can obtain the correlation function for the electric field coefficients, following the methods of Refs.~\cite{Clerk:2008tlb,Foster:2017hbq, Foster:2020fln}. We first separate Eq.~\eqref{EFC frequency} into real and imaginary parts,
\begin{align}
    {\tilde R}(\omega)=\text{Re}[\tilde\alpha(\omega)]\,,\\
    {\tilde I} (\omega)=\text{Im}[\tilde\alpha(\omega)]\,.
\end{align}

The power spectral density (PSD) of a single cavity is given by the sum of their squares $ {\tilde R}^2(\omega)+{\tilde I}^2(\omega) $. Our goal is to generalize such expressions to a statistical treatment of C$a$B detection across $\mathcal N$ detectors. We assume the $i$-th cavity is centered at ${\mathbf x}_i$, where $i = 1, 2, \ldots, \mathcal{N}$. 

The background noise is sculpted by the cavity resonance. We parametrize its size as a product of the transfer function and a Gaussian random variable $b^{(i)}$, which is uncorrelated between the cavities. The resulting dataset is given by:
\begin{align}
\label{Ri}
    {\tilde R}^{(i)}&=\frac{1}{\sqrt T}{\cal T} (\omega)\int^{T/2}_{-T/2} dt \left[b^{(i)}+ g_{a\gamma\gamma}B_0 \sum_{\mathbf k}\int d^3 x_if_{\mathbf k}\cos{\psi^{(i)}_{\mathbf k}}\mathbf{\hat{z}}\cdot\mathbf{e}^*({\mathbf{x}_i})\right]c_{\omega t}\,, 
\end{align}
with $\tilde {I}^{(i)}$ given by the same expression upon taking $c_{\omega t } \rightarrow - s_{\omega t}$.  Here, $ \psi^{(i)}_\mathbf k=\omega_{\mathbf k}t-\mathbf{k}\cdot {\mathbf x}_i+\phi_{\mathbf k}$, $f_{\mathbf k}\equiv\alpha_{\mathbf k}\sqrt{\frac{f(\mathbf k) d^3k}{(2\pi)^3\omega_{\mathbf k}}}$, $c_{\omega t}\equiv\cos{(\omega t)}$, $s_{\omega t}\equiv\sin{(\omega t)}$, and $\mathbf{e}^*({\mathbf{x}_i})$ denotes the eigenfunction in Eq.~\eqref{eigenfunction} for the coordinates ${\mathbf x}_i$. As cavities' size are $\mathcal O(0.1~\mathrm{m})$ and $T$ is sufficiently large, there is no need to consider the averaging out of the spatial coherence.

The background noise satisfies
\begin{align}
\label{background from covariance}
\big<b^{(i)}b^{*(j)}\big>
&=\frac{\delta_{ij} }{Q}\frac{8T^{(i)}_s}{T\omega_0}|\mathcal{T}(\omega)|^2\,,
\end{align}
where $T^{(i)}_s$ refers to the system temperature of the $i$-th cavity~\cite{SHANHE:2023kxz}. In this expression, we approximated $\omega \simeq \omega_0$ for the factor outside the transfer function. This approximation is suitable as long as the cavity is of sufficiently high quality such that the sensitivity is dominated by the narrow range of frequencies around $\omega \simeq \omega_0$. The implication of this is that the coherence length of the filtered axion field is shorter than that of the original field. Furthermore, the coherence time and the spatial correlation of the axion signal is determined by $Q$, not the axion quality factor $Q_a$, which enables us to have a coherent signal across cavity even if the original axion field itself is incoherent. We will always employ this approximation going forward. It follows from Eq.~\eqref{background from covariance} that the two-point functions of the real and imaginary parts of the electric field coefficients are equal.

We now calculate the signal contribution. We further simplify the analysis by taking the axions to be ultra-relativistic such that $|\mathbf k| \simeq \omega_{\mathbf k}$. The desired two-point function of the real components is:
\begin{align}
\label{R_kR_k}
\big\langle {\tilde R}^{(i)}{\tilde R}^{*(j)}\big\rangle=& \frac{(g_{a\gamma\gamma}B_0) ^2}{T}|{\cal T}(\omega)|^2\int^{T/2}_{-T/2}  dt dt'c_{\omega_0 t} c_{\omega_0 t'} \notag\\ &\sum_{\mathbf k,{\bf k}'}\int d^3 x_i d^3 x_j \big<f_{\mathbf k} f_{\mathbf k'}\cos{\psi^{(i)}_{\mathbf k}} \cos{\psi^{'(j)}_{\mathbf k'}}\big>\mathbf{\hat{z}}\cdot\mathbf{e}^*({\mathbf{x}_i})\mathbf{\hat{z}}\cdot\mathbf{e}({\mathbf{x}_j}).
\end{align}
The expectation values of the phase functions vanish unless $\mathbf k = \mathbf k'$. In particular, 
\begin{align}
\label{<coscos>}
    \big\langle\cos\psi_{\mathbf k}^{(i)} \cos\psi^{'(j)}_{\mathbf k'}\big\rangle = \frac{\delta _{{\bf k}{\bf k}'}}{2} \left[\cos\left( \omega_{\mathbf k} (t-t') \right) \cos\left( \mathbf{k}\cdot \mathbf{x}_{ij} \right)+ \sin\left( \omega_{\mathbf k}(t-t')\right) \sin\left({\mathbf k} \cdot \mathbf{x}_{ij}\right)\right]\,,
\end{align}
where $\mathbf{x}_{ij}= \mathbf{x}_i-\mathbf{x}_j$. The amplitudes obey $\langle\alpha_{\mathbf k}^2\rangle=2$ such that $\langle f_{\mathbf k}^2\rangle=2f(\omega_{\mathbf k})(\Delta \omega)^3/ \left[(2\pi)^3\omega_{\mathbf k}\right]$ \cite{Foster:2020fln}. 

In summary, the real-part signal contribution of the covariance matrix element is~\footnote{\begin{minipage}[t]{\linewidth}\raggedright
To obtain Eq.~\eqref{eq:RRij} we used several orthogonal relations in the time integral in the large $T$ limit: 
$$
\int dt\,dt'\,c_{\omega_0 t} c_{\omega_0 t'} \cos\left(\omega_{\mathbf k}(t-t')\right) = \frac{T\pi}{2}\delta(\omega_0 - \omega_{\mathbf k}),\quad
\int dt\,dt'\,c_{\omega_0 t} s_{\omega_0 t'} \cos\left(\omega_{\mathbf k}(t-t')\right) = 0\,,
$$
$$\int dt\,dt'\,s_{\omega_0 t} s_{\omega_0 t'} \cos\left(\omega_{\mathbf k}(t-t')\right) = \frac{T\pi}{2}\delta(\omega_0 - \omega_{\mathbf k}),\quad
\int dt\,dt'\,s_{\omega_0 t} c_{\omega_0 t'} \cos\left(\omega_{\mathbf k}(t-t')\right) = 0\,.
$$
\end{minipage}}
\begin{align}
 \big\langle {\tilde R}^{(i)}{\tilde R}^{*(j)}\big\rangle=\frac{(g_{a\gamma\gamma}B_0)^2\pi}{2(2\pi)^3} \int d\mathbf{\hat n}f(\omega_0)\omega_0 |\mathcal{T}(\omega)|^2 \int d^3 x_id^3 x_j\cos\left(\omega_0\mathbf{\hat n}\cdot\mathbf{x}_{ij}\right)\mathbf{\hat{z}}\cdot\mathbf{e}^*({\mathbf{x}_i})\mathbf{\hat{z}}\cdot\mathbf{e}({\mathbf{x}_j}). \label{eq:RRij}
\end{align}
Note that terms involving arising from the sine function in Eq.~\eqref{<coscos>} vanish and we approximate the summation over $\mathbf k$ with an integral with the angular integral denoting an isotropic distribution of $f(\omega_0)$ with the direction vector $\mathbf{\hat n}$.

Since the imaginary component is equivalent to that of the real component upon sending $c_{\omega_0  t} \rightarrow - s_{\omega_0 t} $ and sine and cosine obey identical orthogonality relationships, $\big\langle {\tilde I}^{(i)}{\tilde I}^{*(j)}\big\rangle$ is equivalent to that of the reals (Eq.~\eqref{eq:RRij}). Additionally, cross-terms between the real and imaginary parts to vanish due to the orthogonality relationships between $c_{\omega_0 t} s_{\omega_0 t}$. Since the signal is uncorrelated with the background noise, the cross-term between the signal and the background vanishes. 

Altogether, the two-point functions of the data between two cavities are: 
\begin{align}
 \big<{\tilde R}^{(i)} {\tilde R}^{*(j) }\big>&= \big<{\tilde I}^{(i)} {\tilde I}^{*(j) }\big> = \left[\frac{4T^{(i)}_s}{Q\omega_0}\delta_{ij} +\frac{\pi (g_{a\gamma\gamma}B_0)^2V\omega_0}{2(2\pi)^3}f(\omega_0)\mathcal{F}_{ij}\right] |\mathcal{T}(\omega)|^2 \label{I_ij(omega)}\,,
 \end{align}
where we introduced a frequency-dependent dimensionless relativistic form factor that encodes the cavity-cavity correlations:
\begin{align}
\mathcal{F}_{ij} &\equiv\frac{1}{V}\int d\mathbf{\hat n}\int d^3 x_id^3x_j\cos{(\omega_0\mathbf{\hat n}\cdot\mathbf{x}_{ij})}\mathbf{\hat{z}}\cdot\mathbf{e}^*({\mathbf{x}_i})\mathbf{\hat{z}}\cdot\mathbf{e}({\mathbf{x}_j}).
\end{align}
We compute its value explicitly in various limiting configurations in the following section.
 
As the transfer function $\mathcal{T}(\omega)$ is included in both the signal and the noise, it cancels in the signal-to-noise ratio. The primary function of the cavity resonance is hence to suppress the background amplitude by a factor of $1/Q$.

\section{Correlation Function Evaluation}
\label{sect: correlation function}

\begin{figure}[t]
\centering
\begin{tikzpicture}
\node[cylinder, shape border rotate=90, draw, minimum height=3cm, minimum width=2cm] (C1) at (0,0) {};
\draw[->, thick] (1.5,2.25) -- (3,2.25) node[right] {$x$};
\draw[->, thick] (2.25,1.5) -- (2.25,3) node[above] {$z$};
\draw[dashed] (1cm, -1.3cm) arc[start angle=0, end angle=180, x radius=1cm, y radius=0.1cm];

\node[cylinder, shape border rotate=90, draw, minimum height=3cm, minimum width=2cm] (C2) at (5,1) {};
\draw[dashed] (6cm, -0.3cm) arc[start angle=0, end angle=180, x radius=1cm, y radius=0.1cm];

\draw[-latex] (C1.bottom) -- ++(.75,2.) node[right, xshift=-0.6cm] {$\mathbf{x}_i$};
\draw[-latex] (C1.bottom) -- (C2.bottom) node[midway, yshift=-0.25cm] {$\mathbf{x}_0$};
\draw[-latex] (C1.bottom) -- (4.5, 1.5) node[midway, yshift=0.3cm] {$\mathbf{x}_j$};
\draw[-latex] (C1.bottom) -- ++(0,3) node[above right] {$\hat{\mathbf{z}}$};

\coordinate (A) at (0, 0.5);
\coordinate (B) at (C1.bottom);
\coordinate (C) at (C2.bottom);

\draw pic["$\theta_0$", draw, -, red, thick, angle radius=1cm, angle eccentricity=1.2] {angle = C--B--A};

\draw[<->] (6.2, -0.5) -- (6.2, 2.5) node[midway, right] {$L$};
\draw[<->] (5, -0.5) -- (6, -0.5) node[midway, below] {$R$};
\end{tikzpicture}
\caption{Geometry of the vectors involved in calculating the correlation function between the electric fields produced in two cavities.}
\label{vector redefinition picture}
\end{figure}

Having derived a general expression for the relativistic cavity form factor \(\mathcal{F}_{ij}\), we now evaluate it explicitly. We begin by shifting the variable \(\mathbf{x}_j\) by the cavity separation vector \(\mathbf{x}_0\), i.e., \(\mathbf{x}_j \rightarrow \mathbf{x}_j - \mathbf{x}_0\), which factorizes the volume integrals over the two cavities. Without loss of generality, we choose \(\mathbf{x}_0\) to lie in the \(x\)-\(z\) plane. The cavity form factor then reads:
\begin{align}
\label{eq:Fij_general}
\mathcal{F}_{ij} = \frac{1}{V} \int d\hat{\mathbf{n}}\, \cos(\omega_0 \hat{\mathbf{n}} \cdot \mathbf{x}_0) \left| \int d^3 x_i\, e^{i \omega_0 \hat{\mathbf{n}} \cdot \mathbf{x}_i} \hat{\mathbf{z}} \cdot \mathbf{e}^* ({\mathbf{x}_i})\right|^2.
\end{align}
Here, \(\mathcal{F}_{ij}\) depends on volume integrals over identical cylindrical cavities of radius \(R\) and height \(L\), and on the integration over the incoming axion direction \(\hat{\mathbf{n}}\).

The inner integral evaluates to:
\begin{align}
\int d^3x \, e^{i \omega \hat{\mathbf{n}} \cdot \mathbf{x}} \hat{\mathbf{z}} \cdot \mathbf{e}^*({\mathbf{x}}) = -i \frac{2\pi R}{\sqrt{V} \omega_0} \frac{1 - e^{i \omega_0 c_\theta L}}{\omega_0 c_\theta} \left[ \frac{J_0(\omega_0 s_\theta R)}{c_\theta^2} \right]\,,
\end{align}
where \(\theta\) is the angle between the \(z\)-axis and \(\hat{\mathbf{n}}\), with shorthand \(c_\theta = \cos\theta\) and \(s_\theta = \sin\theta\). Substituting into Eq.~\eqref{eq:Fij_general} yields
\begin{align}
\mathcal{F}_{ij} 
= \frac{2(2\pi R)^2}{V^2 \omega_0^2} \int_{-1}^1 dc_\theta \int_0^{2\pi} d\phi \, \cos\big[ \omega_0 x_0 (s_\theta c_\phi s_{\theta_0} + c_\theta c_{\theta_0}) \big] 
\frac{1 - \cos(\omega_0 c_\theta L)}{(\omega_0 c_\theta)^2} \left[ \frac{J_0(j_{10} s_\theta)}{c_\theta^2} \right]^2\,,
\end{align}
where \(\theta_0\) is the polar angle of \(\mathbf{x}_0\) (see Figure~\ref{vector redefinition picture}). The \(\phi\) integral can be performed, giving:
\begin{align}
\mathcal{F}_{ij} 
= \frac{4\pi (2\pi R)^2}{V^2 \omega_0^2} \int_{-1}^1 dc_\theta \, \cos(\omega_0 x_0 c_\theta c_{\theta_0}) J_0(\omega_0 x_0 s_\theta s_{\theta_0}) \frac{1 - \cos(\omega_0 c_\theta L)}{(\omega_0 c_\theta)^2} \left[ \frac{J_0(j_{10} s_\theta)}{c_\theta^2} \right]^2\,.
\end{align}

To further evaluate this integral, we expand each angular-dependent factor in a Taylor series. Since \(\omega_0 x_0\) and \(\omega_0 L\) are not generically small, the expansions are not individually convergent, except for the factor \(J_0(j_{10} s_\theta)/c_\theta^2\), which is independent of cavity size or position and can be truncated at \(\mathcal{O}(c_\theta^2)\).

Collecting terms, the resulting expression for \(\mathcal{F}_{ij}\) is
\begin{align}
\label{eq:Fij_series}
\mathcal{F}_{ij} \simeq &\frac{4\pi (2\pi R)^2}{V^2 \omega_0^2} \int_{-1}^1 dc_\theta \,
\left[\sum_{m=0}^\infty \frac{(-1)^m}{(2m)!} (\omega_0 x_0 c_\theta c_{\theta_0})^{2m} \right]
\left[ \sum_{n=0}^\infty \frac{(-1)^n}{(n!)^2 2^{2n}} (\omega_0 x_0 s_{\theta_0})^{2n} (1 - c_\theta^2)^n \right]
 \notag \\
& \times\left[ L^2 \sum_{\ell=0}^\infty \frac{(-1)^\ell}{[2(\ell+1)]!} (\omega_0 L c_\theta)^{2\ell} \right]\left[(1 - \mathcal{J}) c_\theta^2 + \mathcal{J} \right]\,,
\end{align}
where \(\mathcal{J} \equiv (J_1(j_{10}) j_{10}/2)^2 \simeq 0.39\). We employ the identity
\begin{align}
    \int^1_{-1}dc_\theta \,c_\theta^{2\ell}(1-c_\theta^2)^n((1-\mathcal{J})c^2_\theta+\mathcal{J})=\frac{(1+2\ell+2\mathcal{J}(1+n))n!\,\Gamma(\frac{1}{2}+\ell)}{2\Gamma(\frac{5}{2}+n+\ell)}\,.
\end{align}
Substituting it into Eq.~\eqref{eq:Fij_series} and defining the dimensionless variables \(\chi \equiv \omega_0 x_0\) and \(\lambda \equiv \omega_0 L\) gives
\begin{align}
\label{long integral}
    \mathcal{F}_{ij}=\frac{16\pi}{j^2_{10}}\sum_{\{\ell, m, n\}}^\infty\frac{\left[1 + 2(\mathcal{J}(1+n)+\ell+m)\right]c_{\theta_0}^{2m}\,s_{\theta_0}^{2n} \,\chi^{2(m+n)} \, \lambda^{2\ell} \, \Gamma\left(\frac{1}{2} + \ell + m\right)}
{(-1)^{\ell + m + n}\,2^{2n+1}\,(2m)! \,n!\,\Gamma(3 + 2\ell) \, \Gamma\left(\frac{5}{2} + \ell + m + n\right)}\,.
\end{align}
Below, we evaluate Eq.~$\eqref{long integral}$ in various limits.

\begin{figure}[t]
\centering
\includegraphics[width=1\linewidth]{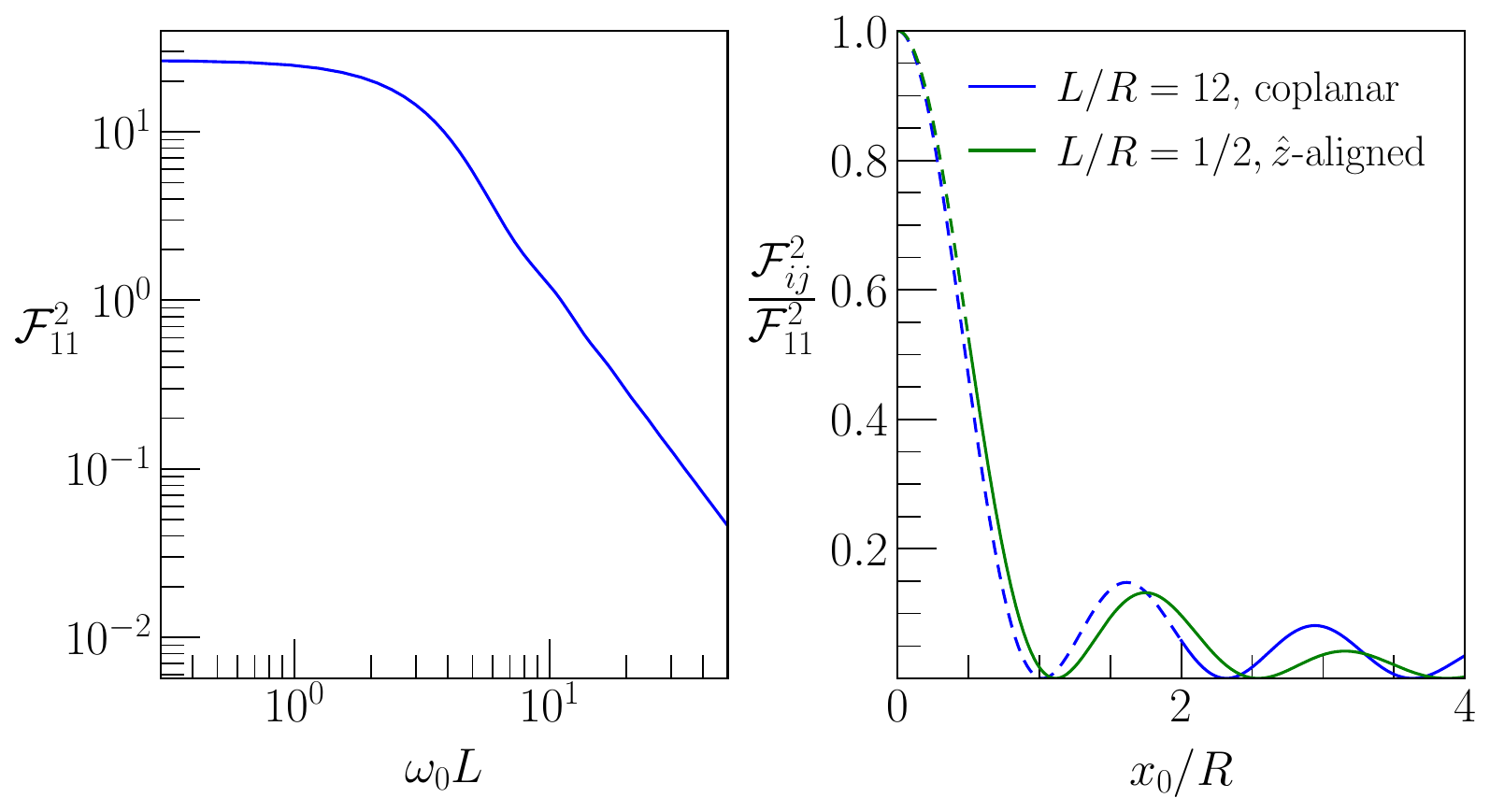}
\caption{
{\bf Left:} Self-correlation \(\mathcal{F}_{ii}\) as a function of \(\omega_0L\), illustrating the limiting behavior of Eq.~\eqref{eq:self_correlation_limits}. {\bf Right:}  Ratio of the inter-cavity correlation (\(i \neq j\)) to the self-correlation (\(i = j\)) as a function of separation normalized by the radius, for \(L/R = 1/2\) (green line) and \(L/R = 12\) (blue line). Blue curves correspond to coplanar cavities; green curves correspond to vertically aligned cavities. Dashed lines indicate unphysical overlapping cavity configurations. Both cases show oscillatory decay with increasing separation, with correlation potentially vanishing at specific distances. 
}
\label{fig: 2 2z}
\end{figure}

\paragraph{Self-Correlation:}  The cavity self-correlation, which is going to be related to the diagonal elements of the covariance matrix below, is obtained by fixing $m = n = 0$ and performing the remaining sum:
\begin{align}
\label{eq:self_correlation}
\mathcal{F}_{ii} = \frac{32 \pi}{j_{10}^2} \frac{1}{\lambda^3} \left[ \lambda - 2 \mathcal{J} \lambda + \mathcal{J} \lambda \cos \lambda + (\mathcal{J} - 1) \sin \lambda + \mathcal{J} \lambda^2 \operatorname{Si}(\lambda) \right]\,,
\end{align}
where \(\operatorname{Si}(x) \equiv \int_0^x \frac{\sin t}{t} dt\) is the sine integral. Its limiting behaviors are (also shown in Figure~\ref{fig: 2 2z} left panel):
\begin{align}
\label{eq:self_correlation_limits}
\mathcal{F}_{ii} \rightarrow
\begin{cases}
\displaystyle \frac{16 \pi (1 + 2 \mathcal{J})}{3 j_{10}^2} & \lambda \ll 1 \\[2ex]
\displaystyle \frac{16 \pi^2 \mathcal{J}}{\lambda j_{10}^2} & \lambda \gg 1
\end{cases}\,.
\end{align}

The fact that the relativistic cavity form factor falls with $\lambda$ (and hence the cavity height $L$) is a generic property unique to relativistic axion detection that tends to prohibit its sensitivity for cavities with $L\gg R $. This was similarly observed in Ref.~\cite{Dror:2021nyr} with a more simplistic treatment of the axion field. This limitation is partially mitigated in realistic experimental designs that typically employ tuning rods. In practice, one would need to recalculate these form factors in the presence of the tuning rods.  

\paragraph{Fat-Cavity Limit:} When \(\lambda \to 0\) (i.e., \(L \ll R / j_{10}\)), the inter-cavity correlation simplifies. Although this limit does not apply to existing experiments, it provides a useful theoretical test case for multi-cavity C$a$B detection and tends to maximize sensitivity. In this limit, the $\ell $ sum in Eq.~\eqref{long integral} converges rapidly, and only the $\ell=0$ term is needed. The result, valid for all \(\theta_0\) and \(x_0\), is
\begin{align}
\label{eq:fat_cavity}
\mathcal{F}_{ij} = \frac{8 \pi}{j_{10}^2} \frac{1}{\chi^3} \left[
(\mathcal{J} - 1) \left( \sin \chi - (1 + 3 c_{2\theta_0}) \chi \cos \chi - c_{2\theta_0} \sin \chi (\chi^2 - 3) \right) + (\mathcal{J} + 1) \chi^2 \sin \chi
\right]\,.
\end{align}
In the case of coplanar cavities ($\theta_0 = \pi/2$), it further reduces to
\begin{align}
\label{simple cavity}
    \mathcal{F}_{ij}= \frac{16\pi}{j^2_{10}}\frac{(\mathcal{J}-1) \chi \cos{\chi} + \sin{\chi} + \mathcal{J} (\chi^2-1) \sin{\chi}}
{\chi^3}\,.
\end{align}

\paragraph{Vertically-Separated Cavities:} For cavities aligned along the vertical axis (\(\theta_0 = 0\)), $n = 0$ and the form factor reduces to
\begin{align}
\label{eq:vertical_cavities}
\mathcal{F}_{ij} &= \frac{16 \pi}{j_{10}^2} \frac{1}{\chi (\chi - \lambda) \lambda^2 (\chi + \lambda)} \Bigg[
2 \chi \cos \lambda \left( \mathcal{J} (\chi^2 - \lambda^2) \cos \chi + (\mathcal{J} - 1) \chi \sin \chi \right) \notag \\
&\quad - 2 (\mathcal{J} - 1) \chi \lambda \cos \chi \sin \lambda + (\chi^2 - \lambda^2) \Big( 2 \sin \chi - 2 \mathcal{J} (\chi \cos \chi + \sin \chi) \notag \\
&\quad + \mathcal{J} \chi \big( -2 \chi \operatorname{Si}(\chi) + (\chi - \lambda) \operatorname{Si}(\chi - \lambda) + (\chi + \lambda) \operatorname{Si}(\chi + \lambda) \big) \Big)
\Bigg]\,.
\end{align}

When two cavities are stacked right on top of each other ($\chi = \lambda$), it becomes
\begin{align}
    \mathcal{F}_{ij}= &\frac{8\pi}{j^2_{10}}\frac{1}{\lambda^3}\Bigg[
-4 \mathcal{J} \lambda \cos\lambda  
+ 2 \mathcal{J} \lambda \cos2\lambda
- 4 (\mathcal{J}-1 ) \sin\lambda\notag\\
&+ (\mathcal{J}-1) \sin2\lambda
+ 2 \lambda \Big(2\mathcal{J}-1
+ 2 \mathcal{J} \lambda (-\text{Si}(\lambda) + \text{Si}(2\lambda))\Big)\Bigg]\,.
\end{align}

The limiting forms of Eq.~\eqref{eq:vertical_cavities} are
\begin{align}
\label{eq:vertical_limits}
\mathcal{F}_{ij} \rightarrow
\begin{cases}
\displaystyle \frac{16 \pi}{j_{10}^2 \chi^3} \left[ -2 (\mathcal{J} - 1) \chi \cos \chi + (-2 + 2 \mathcal{J} + \chi^2) \sin \chi \right] & \lambda \ll 1 \\[2ex]
\displaystyle \frac{16 \pi^2 \mathcal{J}}{\lambda j_{10}^2} & \lambda \gg 1
\end{cases}\,,
\end{align}
which matches the fat-cavity limit at small \(\lambda\).

\paragraph{Cavities in Plane:}  
For cavities lying in the same plane, we set \( m = 0 \) in Eq.~\eqref{long integral}. Summing over \( n \) yields the form factor as an infinite series:
\begin{align}
\label{eq:coplanar_cavities}
\mathcal{F}_{ij} = \frac{16 \pi}{j_{10}^2} \sum_{\ell=0}^\infty \frac{(-1)^\ell 2^{\ell + \frac{1}{2}} \lambda^{2\ell}}{\chi^{\ell + \frac{3}{2}}} \left[ (1 + 2\mathcal{J} + 2\ell)\, J_{\ell + \frac{3}{2}}(\chi) - \mathcal{J}\chi\, J_{\ell + \frac{5}{2}}(\chi) \right] \frac{\Gamma\left(\ell + \frac{1}{2}\right)}{\Gamma(3 + 2\ell)}\,,
\end{align}
where \( \Gamma \) denotes the Gamma function. This expression cannot be fully simplified in closed form, but the sum converges rapidly when \( L / R = \mathcal{O}(1) \), allowing accurate numerical evaluation with a modest number of terms. When \( L / R \gg 1 \), convergence becomes slow, yet the form factor is exponentially suppressed and can be neglected in practice.

In the right panel of Figure~\ref{fig: 2 2z}, we plot the correlation-to-self-correlation ratio as a function of the cavity separation, showing results for coplanar cavities (blue) and cavities stacked vertically (green). Dashed lines show unphysical regions where cavities are overlapping. In the case of coplanar cavities, we fix $L/R=12$ to match the cavity dimensions of the existing ADMX four-cavity array search. In the case of $\hat z$-aligned cavities, we choose a different $L/R$ since otherwise the cavity form factor is negligible in the physical region.

\section{Statistics of a Multi-Cavity Array}
\label{The statistics of a multi-cavity array}

We now use the two-point cavity correlation function derived in the previous section to construct a likelihood function for a prototypical analysis. With this likelihood, we estimate the experimental sensitivity of various cavity array configurations using the Asimov dataset approach. In this method, the dataset of a prospective experiment is taken to be the expectation value of the square root of the variance of the background noise~\cite{Cowan:2010js}, enabling projection of upper limits on signal parameters without requiring Monte Carlo simulations.

From $\mathcal{N}$ detectors, the real and imaginary parts of the Fourier-transformed electric field can be organized into a $2\mathcal{N}$-dimensional data vector:
\begin{equation}
\mathbf{d}_k = [\tilde R_k^{(1)}, \tilde I_k^{(1)}, ..., \tilde R_k^{(\mathcal N)}, \tilde I_k^{(\mathcal N)}]^\intercal\,.
\end{equation}
As discussed in Section~\ref{sec2.3}, cross-terms mixing real and imaginary components vanish, while the purely real and imaginary covariances are equal. Consequently, the $(i, j)$-th block of the $2\mathcal N \times 2\mathcal N$ covariance matrix $\Sigma_k$ takes the form:
\begin{align}
\label{covariance matrix}
\Sigma^{ij}_k = \langle{\tilde R_k}^{(i)}{\tilde R_k}^{*(j)}\rangle\mathbf{I}_2\,,
\end{align}
where $k$ indexes the discrete Fourier modes, and $\mathbf{I}_2$ is the $2\times2$ identity matrix. 

The likelihood function constructed from the covariance matrix is given by
\begin{align}
\mathcal{L}({\mathbf{d}_k}|\boldsymbol{\theta}) = \prod_{\omega_0} \prod_{k=0}^{N-1}
\frac{\exp\left[-\frac{1}{2} \mathbf{d}_k^\intercal \Sigma_k^{-1}(\boldsymbol{\theta}) \mathbf{d}_k\right]}
{\sqrt{(2\pi)^{2\mathcal{N}} |\Sigma_k(\boldsymbol{\theta})|}}\,,
\end{align}
where $|\Sigma_k(\boldsymbol{\theta})|$ denotes the determinant of the covariance matrix, and $\boldsymbol{\theta}$ represents the set of model parameters. Each likelihood term depends implicitly on both $\omega_0$ and $k$, encoding correlations between neighboring frequency bins. The outer product runs over all resonant frequencies $\omega_0$ probed by the experiment, while the inner product index $k$ is determined by the total observation time and sampling interval. In the limit of dense time sampling, this discretization approaches the continuous frequency variable $\omega = 2\pi k/T$.

The covariance matrix includes both nuisance parameters, such as the system temperature, and signal parameters, such as $g_{a\gamma\gamma}$ and the axion phase-space distribution. In a frequentist analysis, the nuisance parameters are profiled over, while in a Bayesian framework, they are marginalized using priors. For the purpose of projections, these approaches yield comparable results; we adopt the simpler strategy of fixing nuisance parameters to benchmark values.

We define the log-likelihood ratio test statistic (TS) as:
\begin{equation}
\label{asimov}
q(\boldsymbol{\theta}) = 2\left[\ln \mathcal{L}_{S+B}(\{\mathbf{d}_k\}|g_{a\gamma\gamma}, f(\omega), T_s, \cdots) - \ln \mathcal{L}_B(\{\mathbf{d}_k\}|T_s, \cdots)\right]\,,
\end{equation}
where $S+B$ and $B$ refer to signal plus background and background-only hypotheses, respectively. We now write the total covariance as a sum of signal and background components:
\begin{align}
\Sigma_k = S_k + B_k\,,
\end{align}
where $B_k$ is diagonal as given in Eq.~\eqref{background from covariance}. In the small-signal limit, the Asimov expectation of the test statistic becomes:
\begin{align}
\label{teststatistics}
\langle q\rangle = -\frac{1}{2} \sum_{\omega_0} \sum^{N-1}_{k=1} \text{Tr}\left[S_k B_k^{-1} S_k B_k^{-1} \right]\,.
\end{align}

In a simplified scenario where both the signal and background are independent of $\omega_0$ and $k$, the test statistic scales with the number of bins. The summations over $\omega_0$ and $k$ can then be approximated by their respective bin counts. For $k$, the signal bandwidth may in principle extend to frequencies $\omega \gg \omega_0 (1 + 1/Q)$, since both the signal and noise scale with the transfer function. Nevertheless, we restrict the integration range to the resonant bandwidth, as we expect additional noise sources to dominate off resonance. 

The number of \( \omega_0 \) bins is determined by dividing the cavity bandwidth, \( 2\omega_0/Q \), into frequency intervals of width \( 2\pi/T \), yielding \( N \simeq \omega_0 T / \pi Q \). More precisely, we carry out the resonant frequency sum approximately it as an integral, using a frequency step size $\omega_0/Q$:
\begin{equation}
\sum_{\omega_0} \rightarrow \int \frac{Q}{\omega_0} \, d\omega_0\,.
\end{equation}
Typical resonant cavity experiments do not employ simple cylindrical cavities, but instead introduce tuning rods that alter the resonant frequency. We do not take this into account in our study. Instead, we assume cavity radius changes with $L$ and $x_0$ fixed. Thus, in projecting sensitivity, we use the cylindrical cavity relation $R = j_{10}/\omega_0$, and sum over all $ \omega _0 $ bins.

To make projections, we must assume a specific model for the axion frequency spectrum. To this end, we assume an isotropic power-law form for the axion relic energy density per unit log frequency: $\Omega_a(\omega) = \Omega_0 (\omega/\omega_*)^\beta$, with reference scale $\omega_*$. For concreteness, we assume $ \omega _\ast $ takes a value near the center of ADMX’s sensitivity peak $\omega_* = 2.7~\mathrm{ \mu eV}$ and that the power-law frequency spectrum persists over the approximate frequency range scanned by ADMX 1.9--3.53~$\rm{\mu eV}$~\cite{ADMX:2003rdr, ADMX:2009iij}. This corresponds to a phase-space distribution:
\begin{align}
f(\omega) = \frac{2\pi^2 \rho_c}{\omega_*^4} \, \Omega_0 \left( \frac{\omega}{\omega_*} \right)^{\beta - 4}\,,
\end{align}
where $\rho_c = 3 M_{\text{Pl}}^2 H_0^2$ is the critical density, written here in terms of the Planck mass and Hubble constant. 

Our goal is to set a projected 95\% upper limit on $\Omega_0$, assuming no true signal. Accordingly, we set $\Omega_0 = 0$ in the true model and use $q = -2.71$ for the 95\% confidence level projection. A model is then specified by three parameters $ \Omega _0 , \beta , g _{ a \gamma \gamma } $. We find that all array configurations studied show qualitatively similar behavior as $\beta$ varies. In particular, for fixed $g _{ a \gamma \gamma } $, the limit on $\Omega_0$ is maximized at one point and decreases as $ \beta $ moves away the maximum. For projections, we fix $\beta = 0$. Depending on the cavity geometry, varying $\beta$ within the range $-3 < \beta < 3$ results in a monotonic increase in $\Omega_0$, with the value at $\beta = 3$ approximately four times larger than at $\beta = -3$.

Sensitivity depends on the relativistic form factor integrated over the bandwidth of scanned resonant frequencies from $\omega_{\rm{min}}$ to $\omega_{\rm{max}}$:
\begin{equation}
I_{ij} \equiv \int_{\omega_{\rm{min}}}^{\omega_{\rm{max}}} \frac{d\omega_0}{\omega_*} \left( \frac{\omega_0}{\omega_*} \right)^{2\beta - 8} \mathcal{F}_{ij}^2\,.
\label{eq:Iij}
\end{equation}
This integral cannot be performed analytically in general, and we evaluate it numerically in formulating our projections. The cavity-averaged form factor integral is defined as
\begin{equation}
\label{cavity-averaged form factor integral}
\bar{I} \equiv \frac{1}{\mathcal{N}} \sum_{ij} I_{ij}\,.
\end{equation}
In a fully correlated array, $\bar{I}$ scales linearly with $\mathcal{N}$, while in an uncorrelated array, $\bar{I}$ remains constant as the array size is increased. The former only applies for very specific arrays where $\mathcal{F}_{ii}\sim \mathcal{F}_{ij}$, while for most of the other arrays, $\mathcal{F}_{ii}$ is larger and $\bar{I}$ remains close to a constant.

The resulting upper limit on $\Omega_0$ for $\beta = 0$ and $\mathcal{N}$ identical cavities is:
\begin{align}
\label{upper limit expression}
\Omega_0 \simeq 1.1 \times 10^{-3} 
\left[\frac{7~\mathrm{T}}{B_0}\right]^2
\bigg[\frac{g^{\mathrm{SE}}_{a\gamma\gamma}}{g_{a\gamma\gamma}}\bigg]^2
\left[\frac{10^5}{Q}\right]
\left[\frac{T_s}{148~\mathrm{mK}}\right]
\sqrt{\frac{100~\mathrm{s}}{T}}
\left[\frac{97~\mathrm{cm}}{L}\right]
\left[\frac{\omega_{\rm{max}}-\omega_{\rm{min}}}{\omega_*}\right]^{3.5}
\sqrt{\frac{4}{\mathcal{N}}}
\sqrt{\frac{1}{\bar{I}}}\,,
\end{align}
where we benchmark parameters against the ADMX experiment~\cite{ADMX:2018gho, Foster:2017hbq}. The coupling $g_{a\gamma\gamma}$ is normalized to the solar and stellar bounds from CAST and horizontal branch stars, with $g^{\mathrm{SE}}_{a\gamma\gamma} = 6.6\times 10^{-11}~\mathrm{GeV}^{-1}$~\cite{CAST:2017uph, Ayala:2014pea, Carenza:2020zil}. We also take the continuum limit of the Fourier transform. 

Beyond experimental parameters, the sensitivity depends on both the number and geometry of the cavities. The scaling with cavity number is explicit in Eq.~\eqref{upper limit expression}. In the fully incoherent limit, the sensitivity on $ \Omega _0 $ scales as $\mathcal{N}^{-0.5}$. In the coherent limit, $ \bar{ I} $ scales with $ {\cal N} $ leading to sensitivity scaling as $\mathcal{N}^{-1}$. In the next section, we explore how the cavity array geometry shapes this behavior. 

\section{Evaluating Upper Limits for Resonant Cavity Arrays}
\label{sec:5 evaluating upper limit}

Having derived the sensitivity expression for resonant cavity arrays targeting the C$a$B, we now optimize the parameters of the array and present projections for a range of array geometries. These include both the current four-cavity configuration of ADMX and its proposed eighteen-cavity upgrade. We consider arrays composed of identical cylindrical cavities and compute their sensitivity using Eqs.~\eqref{eq:Iij} and~\eqref{upper limit expression}, along with the relativistic form factors ${\cal F}_{ij}$ derived in Section~\ref{sect: correlation function}.

\subsection{Geometric Optimization in Multi-Cavity Arrays}
\label{subsec: Geometric Optimization in Multi-Cavity Arrays}

Given Eqs.~\eqref{upper limit expression} and ${\cal F}_{ij}$ in various limits, we can analyze how different parameters affect sensitivity and use these insights to design an optimized multi-cavity array.

Several key features emerge. First, while it may seem that increasing the cavity height \( L \) should enhance sensitivity by increasing the conversion volume, this intuition breaks down at large \( L \). As discussed earlier, the relativistic form factor decreases with increasing \( L \), which suppresses \( \bar{I} \) and cancels the \( L \)-dependence in Eq.~\eqref{upper limit expression} in the large-\( L \) limit. This behavior agrees with the result from Ref.~\cite{ADMX:2023rsk}, which is, caused by the suppression due to multiple oscillations over the cavity.

Second, while it is the case for the vertically stacked array, the best sensitivity is not always achieved at the smallest possible inter-cavity separation. For example, in the case of coplanar array with \( L/R = 12 \)—the aspect ratio relevant for ADMX's existing multi-cavity array in the blue line of Figure~\ref{fig: 2 2z}—the form factor is maximized near \( x_0 \simeq 3R \), rather than at the minimum allowed separation \( x_0 = 2R \). While the precise value of the optimal spacing depends on \( L/R \), this behavior is generic.
\begin{figure}
    \centering
    \includegraphics[width=1\linewidth]{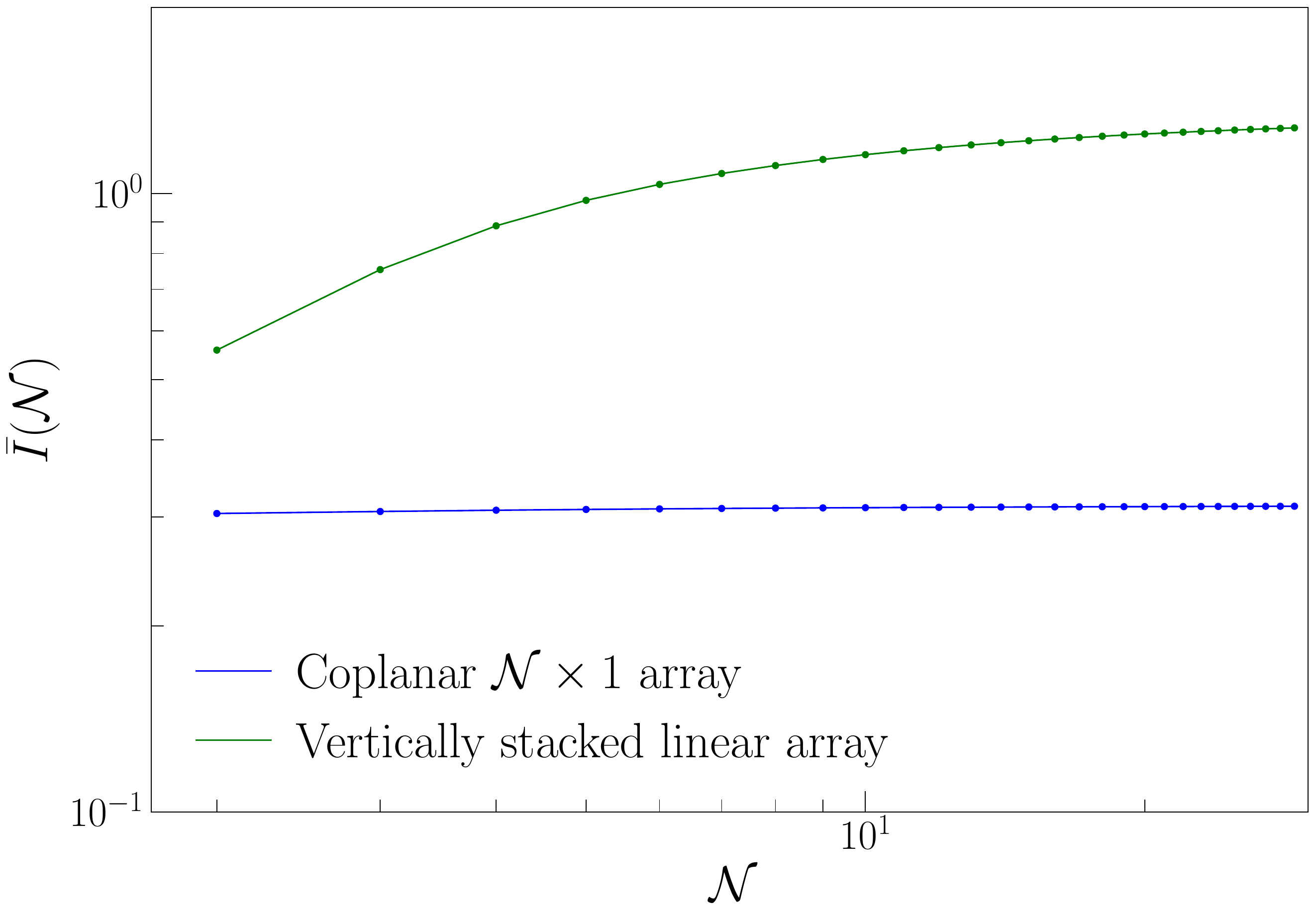}
    \caption{$\bar I(\mathcal N)$ versus $\mathcal N$ for a vertically stacked linear array in Section~\ref{vertically stacked fat cavities} (green) and a coplanar $\mathcal N\times 1$ array (blue). Each cavity has a dimension of $R= 25~\mathrm{cm}$ and $L = 5~\mathrm{cm}$ and is touching the neighboring cavities. The green line shows a coherent configuration with $\bar I(\mathcal N)$ increasing linearly and then approaching a constant. On the other hand, $\bar I(\mathcal N)$ of an incoherent configuration is effectively constant regardless of the size of the array.}
    \label{fig:omega over N}
\end{figure}

Finally, while increasing the number of cavities enhances sensitivity, the scaling behavior of \( \Omega_0 \) with respect to \( \mathcal{N} \) can change in a larger array depending on the configurations. When the length of each cavity is held fixed, the sum of self-correlation terms in \( \bar{I} \) remains independent of \( \mathcal{N} \), while the sum of cross-correlation terms can scale up to \( \mathcal{N} \), depending on the array geometry. This scaling reflects the number of distinct cavity pairs. This manifests that if we are using a highly correlated configuration, such as the one considered in the next subsection (green line in Figure~\ref{fig:omega over N}), cross-correlations can be dominant with respect to self-correlations, and the sensitivity scales as \( \Omega_0 \propto \mathcal{N}^{-1} \) initially. As the array size increases, cross-correlation effect becomes weaker compared to self-correlation as the separation becomes larger. Thus, the effect of increasing the array size diminishes and the scaling of the sensitivity drops to \( \Omega_0 \propto \mathcal{N}^{-1/2}\). However, in an incoherent configuration, such as a coplanar $\mathcal N\times 1$ array in Figure~\ref{fig:omega over N}, cross-correlations are always subdominant and $\bar I$ stays constant and the sensitivity scaling stays as \( \Omega_0 \propto \mathcal{N}^{-1/2}\).

\subsection{Vertically Stacked Linear Arrays}
\label{vertically stacked fat cavities}
We have seen from above that the best sensitivity is achieved by both increasing the number of cavities and optimizing the cavity height \( L \). However, there also exist some geometry-dependent parameters that complicate the optimization, such as the separation and $\mathcal{N}$ in $\bar{I}$.

To utilize the aforementioned effects, we consider an optimized configuration in which a single tall cavity is partitioned into multiple ``fat'' cavities stacked vertically. This approach increases the number of cavities while avoiding the suppression associated with large \( L/R \) ratios. Furthermore, it ensures that the sensitivity is in the coherent limit by taking advantage of the large-valued regime in the green line of Figure~\ref{fig: 2 2z}.

\begin{figure}
    \centering
    \includegraphics[width=1\linewidth]{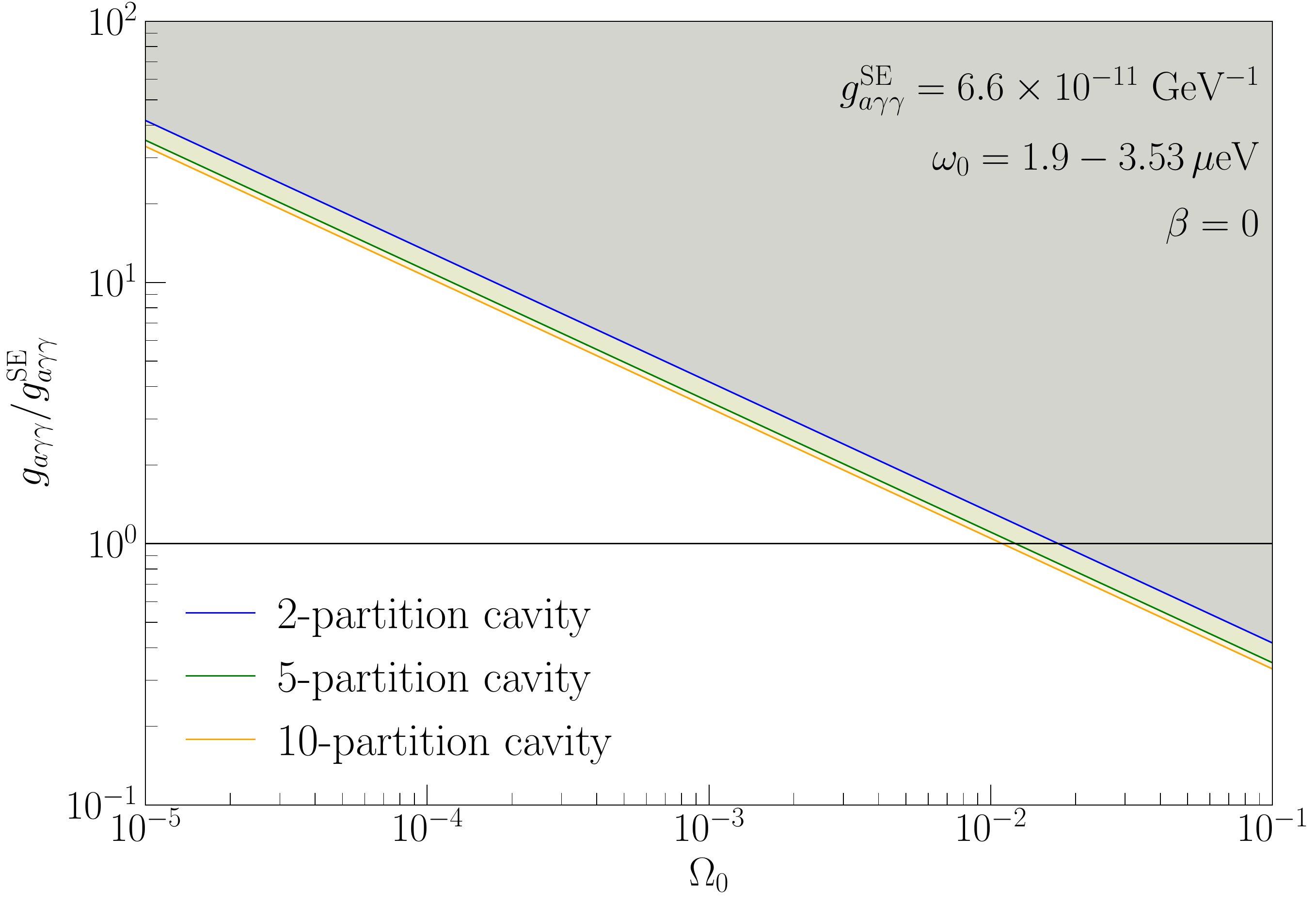}
    \caption{Normalized $g_{a\gamma\gamma}$ (with respect to the CAST limit $g_{a\gamma\gamma}^{\mathrm{SE}}=6.6\times 10^{-11}\,\mathrm{GeV}^{-1}$) versus the 95\% upper limit on $\Omega_0$ for a $1~\mathrm{m}$-tall cavity partitioned into different numbers of stacked segments. All other parameters match those in Section~\ref{The statistics of a multi-cavity array}. The black line shows $g_{a\gamma\gamma}/g_{a\gamma\gamma}^{\mathrm{SE}}=1$, and the shaded regions denote sensitivity bands.}
    \label{fig:fat cavity}
\end{figure}

Figure~\ref{fig:fat cavity} illustrates the resulting sensitivity in the \( g_{a\gamma\gamma} \)--\( \Omega_0 \) plane for vertically stacked configurations. We fix the total height of the experimental setup to \( 1~\mathrm{m} \) and vary the number of partitions \( \mathcal{N} \). All other parameters are held fixed as in Section~\ref{The statistics of a multi-cavity array}. In the coherent limit, sensitivity improves with increasing \( \mathcal{N} \), although the total magnetic field volume remains constant. This leads to sensitivity saturation for sufficiently large \( \mathcal{N} \). In particular, when \( \mathcal{N} \) is large, the sum of cross-correlation terms in Eq.~\eqref{upper limit expression} can dominate the average power \( \bar{I} \), scaling approximately linearly with \( \mathcal{N} \).~\footnote{There is a subtle reason why $\bar I$ scales linearly with $\mathcal{N}$ even for large $\mathcal{N}$ unlike in Section~\ref{subsec: Geometric Optimization in Multi-Cavity Arrays}. Previously, the size of an array is allowed to expand without any constraint while the total length is fixed here. Therefore, even if $\mathcal{N}$ increases, the separation is bounded and the cross-correlations stay dominant.
} Since we hold the total height \( \mathcal{N} L \) fixed, the sensitivity gain saturates for \( \mathcal{N} \gg 1 \).

\subsection{Comparison With Existing Proposals}
\label{sec: Comparison with Existing Proposals}

\begin{figure}
    \centering
    \includegraphics[width=1\linewidth]{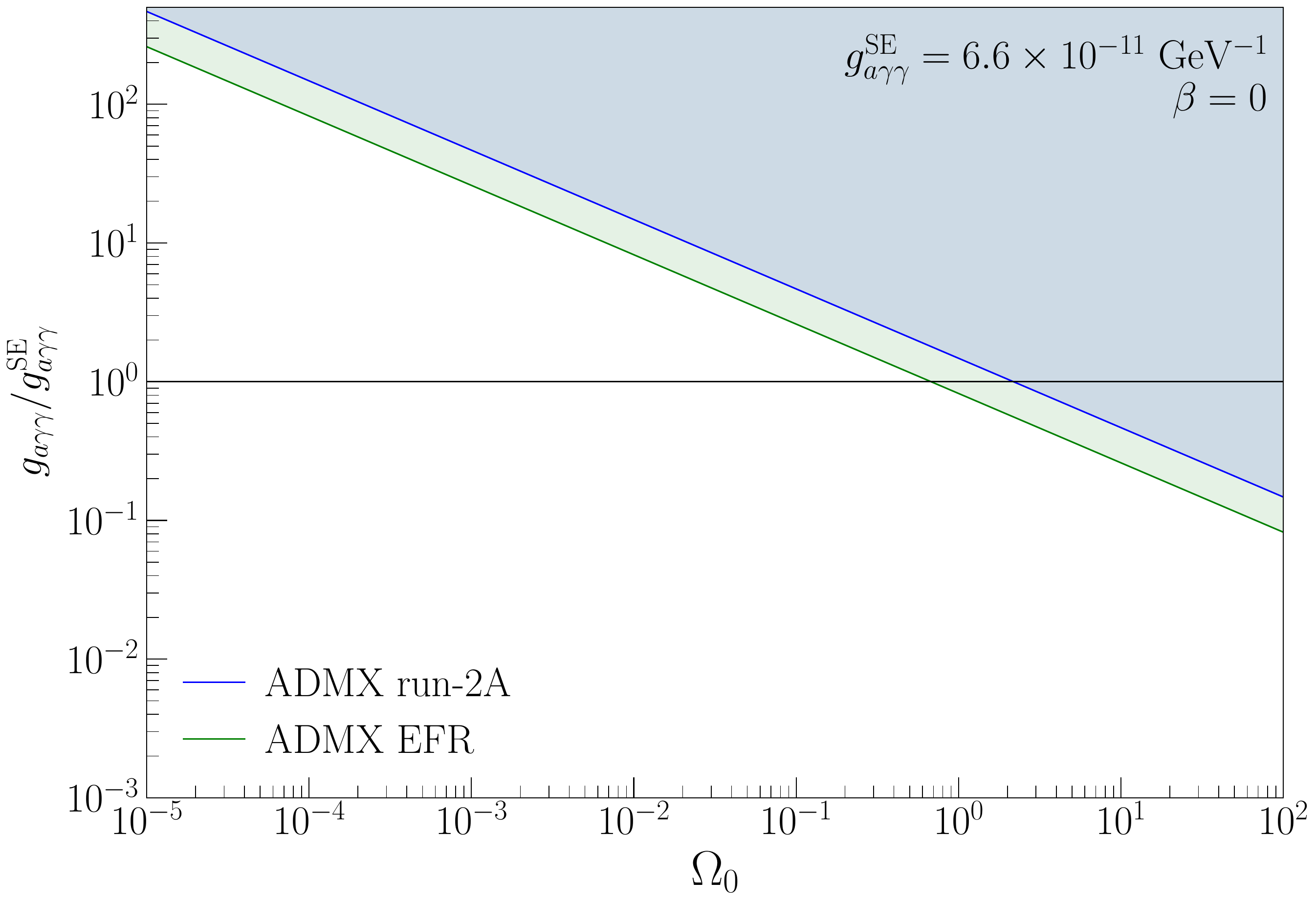}
    \caption{Normalized $g_{a\gamma\gamma}$ (with respect to the CAST limit $g_{a\gamma\gamma}^{\mathrm{SE}}=6.6\times 10^{-11}\,\mathrm{GeV}^{-1}$) versus the 95\% upper limit on $\Omega_0$ for existing and future ADMX proposals. The black line indicates the CAST limit, $g_{a\gamma\gamma}/g_{a\gamma\gamma}^{\mathrm{SE}}=1$. The colored regions denote sensitivity bands. The blue and green lines correspond to ADMX run-2A and ADMX-EFR, respectively. All other parameters are as in Section~\ref{The statistics of a multi-cavity array}, with the magnetic field for ADMX-EFR taken to be 9.4~T and $Q = 1.8\times 10^5$.}
    \label{fig: ADMX proposals}
\end{figure}
ADMX has a current run, ``run-2A'', which targets a tuning range of 1.5--2.2 GHz using a coplanar square four-cavity array with $R = 8~\mathrm{cm}$ and $L = 97~\mathrm{cm}$.  This multi-cavity system maintains frequency lock across cavities to maximize volume utilization within the ADMX magnetic bore~\cite{Yang:2020xsc}. For dark matter searches, the system effectively operates as a single higher-frequency cavity. If repurposed for C$a$B searches, the cavities do not always constructively interfere, reducing sensitivity. Figure~\ref{fig: ADMX proposals} presents its sensitivity projections in the $g_{a\gamma\gamma}$--$\Omega_0$ plane (blue). As mentioned previously, we do not incorporate the tuning rod into our analysis, so our projections should only be considered qualitatively. Instead, we assume each resonant frequency search corresponds to a cylindrical cavity with a differing radius. Since $ R = 8~ {\rm cm} $ is significantly smaller than that we assumed above, the sensitivity is somewhat degraded relative to the idealized setups examined earlier.

The upcoming ADMX-EFR (Extended Frequency Range) project aims to explore the 2--4 GHz band over a three-year operation, employing a more powerful 9.4~T magnetic field, a cavity factor $Q$ of $1.8\times 10^5$, and an 18-cavity array. Each cavity has a radius of 6.4~cm and a height of 1~m, arranged in a hexagonal configuration with a vacant center~\cite{Knirck2023}. Although the cavities do not fit entirely within the volume of the original single cavity, the total effective volume remains comparable. Its sensitivity is also shown in Figure~\ref{fig: ADMX proposals} (green). Even with a smaller radius for each individual cavity, the larger array size, stronger magnetic field, and higher quality factor enhances the prospective sensitivity relative to run-2A. However, since both the current and the future run employ incoherent configurations, the sensitivity scaling due to the number of cavities does not enhance very quickly.

From these investigations, it is evident that improving sensitivity by an order of magnitude through cavity arrangement alone is difficult. In addition, coplanar configuration is not a coherent system; therefore, no change in the scaling with respect to the $\mathcal N$ is present. However, other avenues remain promising. According to Eq.~\eqref{upper limit expression}, the sensitivity limit scales as $\Omega_0 \propto B^{-2}$, $\Omega_0 \propto T_s$, and $\Omega_0 \propto Q^{-1}$. Therefore, optimizing all these parameters may be more effective than simply increasing the number of cavities or changing the geometry.

Of particular interest is the regime where \(\Omega_0 \sim 10^{-5}\). This value is motivated by cosmological considerations: if C$a$B was produced during epochs prior to Big Bang nucleosynthesis (BBN) or recombination, it would be a form of dark radiation in the early universe and a given \(\Omega_0\) today would imply a shift in \(\Delta N_{\mathrm{eff}}\). Observational bounds on \(\Delta N_{\mathrm{eff}}\) from the Planck satellite measurements of the cosmic microwave background $\Delta N_{\rm eff} = 0.34$~\cite{Planck:2018vyg} constrain \(\Omega_0\) to $7\times10^{-6}$ for $\beta = 0 $ and frequency range 1.9--3.53$~\rm{\mu eV}$. Constraints from BBN~\cite{Aver:2015iza, Peimbert:2016bdg, Cooke:2017cwo} also give a similar result. Eq.~\eqref{upper limit expression} tells us that in order to reach \(\Omega_0 \sim 10^{-5}\) using the configuration in Section \ref{vertically stacked fat cavities}, for example, we need to have $B = 20~\mathrm{T}, Q = 10^6, T_s = 47~\mathrm{mK}$, and a measurement time of 1000 seconds, not to mention the frequency lock and the capability of obtaining independent readout across cavities. Consequently, achieving experimental sensitivity at or below this threshold is challenging, but essential for probing cosmologically relevant axion populations.

\section{Conclusion}
The detection of a relativistic axion background is strongly motivated by cosmological considerations. However, existing experiments have not yet developed effective strategies to distinguish its broad spectrum from background noise. A high–$Q$ resonant cavity acts as a narrowband filter with bandwidth $\omega_0/Q$. Thus, the measured axion-induced electric field acquires a spatial correlation length set in part by the larger of the intrinsic axion coherence scale and the cavity quality factor, rather than by the broad axion bandwidth. In other words, even if the C$a$B is weakly correlated over experimental length scales, the filtered field that we read out from the cavity can remain coherent across multiple detectors. This enables us to utilize the cross-correlation between the outputs from multiple nearby detectors, thereby providing an effective means to discriminate between the signal and the background noise.

In this work, we derived the expression for the electric field signal generated inside resonant multi-cavity arrays and constructed the associated correlation function using the statistical properties of the axion field. Within a likelihood framework, we placed projected constraints on model parameters, assuming an isotropic phase space distribution. By incorporating relativistic form factors into the correlation function, we explored a range of experimental configurations and established the relationship between cavity-to-cavity correlations and axion parameters. For concreteness, we assumed a power-law phase space distribution for the axion background, characterized by an energy density per unit logarithmic frequency, $\Omega_0$, with $\Omega_0 \sim 10^{-5}$ corresponding to the level constrained by the CMB and BBN.

We analyzed various cavity geometries and arrangements to enhance signal correlations. In general, compact configurations exhibit stronger correlations, and cavities with $R \gtrsim L $ tend to yield stronger signals. We identified a particularly effective configuration: vertically stacked fat-cavity arrays, for which the sensitivity to $\Omega_0$ is optimal. Of course, further improvements are possible by optimizing not only the array geometry but also the key system parameters such as magnetic field strength, operating temperature, and quality factor. 

Similar correlation-based strategies could also be applied to other axion detection schemes, including those employing nuclear magnetic resonance~\cite{Graham:2013gfa,Budker:2013hfa, Dror:2022xpi} or LC circuits~\cite{Sikivie:2013laa,Chaudhuri:2014dla,Kahn:2016aff}. At the same time, our results point to an important geometric limitation of conventional microwave cavity arrays. In our setups, the correlation length of the filtered axion-induced signal is on the order of the cavity size. As a result, it is difficult to maintain strong correlations between distinct cavities without running into mechanical constraints. Therefore, superconducting radio-frequency (SRF) cavities  with $Q\sim10^{10}$--$10^{12}$ \cite{Janish:2019dpr,Giaccone:2022pke} and LC resonators, for example, are especially interesting in this regard. In those systems the resonance is not fixed purely by geometric mode structure, so it is possible generate long-lived, phase-coherent signals in multiple detectors without being forced into a particular cavity aspect ratio or packing geometry. This could allow more favorable scaling with detector number than what we find for standard cavity arrays. A detailed study of SRF- and LC-based arrays is left for future work. With continued refinement and the inclusion of realistic detector effects, the framework developed here provides a promising path toward probing relativistic axions and, hopefully, discovering their cosmological role.

\acknowledgments
The authors thank Nicholas Rodd and Pierre Sikivie for useful discussions. The research of JD is supported in part by the U.S. Department of Energy grant number DE-SC0025569. The research of SC is supported in part by the Institute of High Energy Physics and Astrophysics 2024 summer research fellowship.

\bibliographystyle{JHEP}
\bibliography{biblio.bib}
\end{document}